\documentclass[aps,prl,twocolumn,superscriptaddress]{revtex4}

\usepackage{psfig}
\begin{document}
\psfigurepath{.:plot:figures}

\title{Novel dynamic scaling regime in hole-doped La$_2$CuO$_4$}

\author{Wei Bao}
\affiliation{Los Alamos National Laboratory, Los Alamos, NM 87545}
\author{Y. Chen}
\affiliation{Los Alamos National Laboratory, Los Alamos, NM 87545}
\author{Y. Qiu}
\affiliation{NIST Center for Neutron Research, National Institute of Standards and Technology, Gaithersburg, MD 20899}
\affiliation{Dept.\ of Materials and Nuclear Engineering, University of
Maryland, College Park, MD 20742}
\author{J.L. Sarrao}
\affiliation{Los Alamos National Laboratory, Los Alamos, NM 87545}

\date{\today}

\begin{abstract}
Only 3\% hole doping by Li is sufficient to suppress the
long-range 3-dimensional (3D) antiferromagnetic order in La$_2$CuO$_4$. 
 The spin dynamics of such a 2D spin liquid state at $T\ll J$ was 
investigated with
measurements of the 
dynamic magnetic structure factor $S(\omega,{\bf q})$, 
using cold neutron spectroscopy,
for single crystalline La$_2$Cu$_{0.94}$Li$_{0.06}$O$_4$.
 $S(\omega,{\bf q})$ peaks sharply at $(\pi,\pi)$ and crosses over around 
50~K from $\omega/T$ scaling to 
a novel low temperature regime characterized by a constant energy scale.  
The possible connection to a crossover from the quantum critical to the 
quantum disordered regime of the 2D antiferromagnetic spin liquid is 
discussed.
\end{abstract}


\maketitle

The experimental investigation of the spin dynamics of 
doped Heisenberg antiferromagnets on a square lattice is 
essential to understanding cuprate superconductors, as 
well as to current research on quantum phase 
transitions\cite{subir_rev}.
For the structurally simplest laminar cuprate, 
La$_2$CuO$_4$, the 3-dimensional (3D)
antiferromagnetic N\'{e}el order due to weak interlayer 
magnetic interaction can be suppressed by 2-3\% hole 
doping using Sr, Ba or Li,
thus allowing experimental investigation of the quantum 
spin dynamics of a 2-dimensional (2D) spin liquid in a
wide temperature range, $0< T \ll J/k_B\sim$ 1000~K\cite{2dheis}.

At the critical doping concentration, $y_c$, of 
La$_2$CuO$_4$ and at $T=0$, 
namely at the quantum critical point, spin dynamics is described
by classical critical dynamics in 2+1 dimensions\cite{2dheis}. 
At a finite temperature, the extra dimension in imaginary time 
is reduced to a finite thickness of $\hbar c/k_B T$, 
where $c$ is the spin wave velocity. 
As a consequence of this long wavelength cutoff, spin dynamics
follows the quantum critical $\hbar \omega/k_B T$ scaling\cite{2dheiqd}.
This has been experimentally observed in Ba and Sr doped La$_2$CuO$_4$ 
in previous
inelastic neutron scattering studies\cite{la2smha,la2keimer,la2gas}.
At $T=0$ and above the critical doping concentration $y>y_c$,
zero-point quantum fluctuations set another long wavelength cutoff. 
Therefore, for $y>y_c$, when the thermal cutoff length, $\hbar c/k_B T$, 
reaches the quantum cutoff during cooling, a crossover in spin dynamics
from the quantum critical (QC) regime, where the energy scale is $k_BT$, to
the quantum disordered (QD) regime, where the energy scale is fixed at infrared
by the quantum cutoff, occurs\cite{2dheis,2dheiqd}.
This crossover, however, has not been explicitly 
investigated experimentally.
Here, we report a cold neutron inelastic scattering study 
on spin dynamics in Li doped La$_2$CuO$_4$ from 150~K to 1.5~K. 
A crossover from the $\omega/T$ scaling 
to a new low temperature regime, which features an infrared energy 
cutoff, is directly observed.

The critical doping concentration, $y_c$, for Li doped 
La$_2$CuO$_4$ is 3\%\cite{Rykov,Li214}. In this study, a $y$=6\% 
single crystal, La$_2$Cu$_{0.94}$Li$_{0.06}$O$_4$,
weighing 2.1 $g$, was grown in
CuO flux, using isotopically enriched $^7$Li (98.4\%) to reduce
neutron absorption of natural Li. 
The crystal has orthorhombic $Cmca$ symmetry with
lattice parameters $a=5.351$~\AA, $b=13.15$~\AA\ and $c=5.386$~\AA\
at 295~K. Using this orthorhombic unit cell to label reciprocal {\bf q}
space, the ($\pi,\pi$) point in the square lattice notation
splits to (100) and (001) points.
Neutron scattering signal is observed
at (100) type but not at (001) type Bragg points, as in the
stoichiometric antiferromagnet La$_2$CuO$_4$\cite{la3dv}.
Measurements of seven independent (100) type Bragg peaks, 
using thermal neutrons
to reach 6~\AA$^{-1}$, confirm the magnetic origin of these peaks.
This conclusion is consistent with previous thermal 
neutron scattering studies
on Li doped La$_2$CuO$_4$\cite{bao99a}, which, however, did not have
sufficient energy resolution to resolve the energy spectra at low temperatures.
Here, we focus on spin dynamics near {\bf q}=(100), 
taking advantage of the enhanced energy 
resolution of the cold neutron triple-axis 
spectrometer SPINS at NIST with fixed $E_f=3.7$~meV
or $E_f=5$~meV. The (002) reflection of pyrolytic graphite was used for both
the monochromator and analyzer.
A cold Be or BeO filter was put before the analyzer to eliminate higher
order neutrons.
Horizontal Soller slits of 80' were used before and after the sample.
The temperature of the sample was regulated by a pumped He 
cryostat.

\begin{figure}
\psfig{file=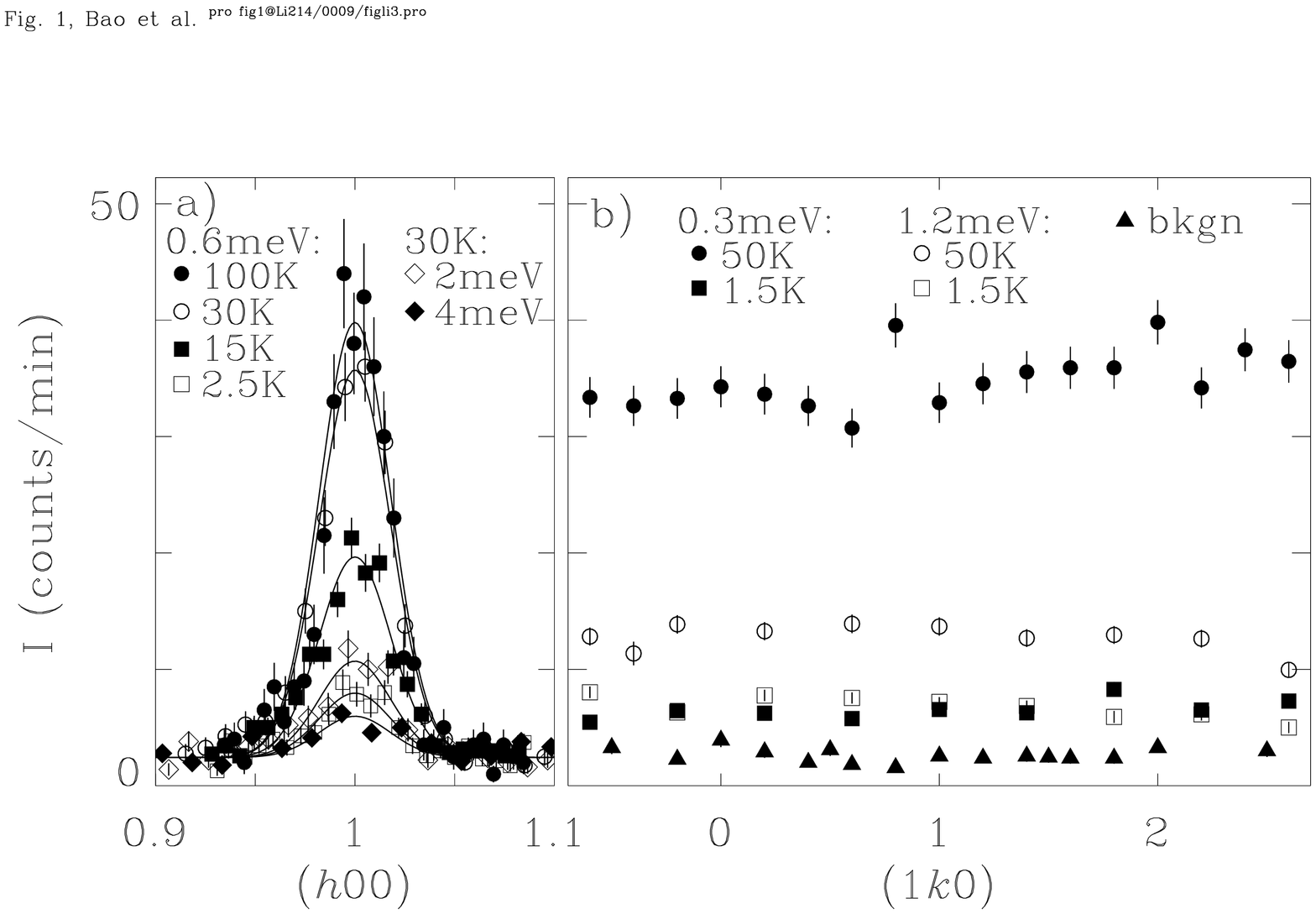,width=.95\columnwidth,angle=90,clip=}
\vskip -1.3cm
\caption{
Const.-$\hbar\omega$ scans with various energy transfers and at various temperatures (a) in the basal CuO$_2$ plane and (b) perpendicular to the
basal plane.
The {\bf q}=(100) here in the orthorhombic notation corresponds to
the ($\pi,\pi$) point of the CuO$_2$ square plane.
Background in (b), triangles, was measured at 1.5~K either at $-0.5$~meV, or
at (1.2,$k$,0) with $\hbar\omega$=0.3 and 1.2~meV.
}
\label{fig1}
\end{figure}

Fig.~\ref{fig1} shows some constant-energy ($\hbar\omega$) scans (a) in the 
basal plane and (b) perpendicular to the basal plane
around {\bf q}=(100) at various energies and temperatures.
The sharp peak at ($\pi,\pi$) in Fig.~\ref{fig1}(a)
does not show appreciable change in its width 
over a wide energy and temperature range, 
suggesting a resolution-limited in-plane peak. 
Flat scans in Fig.~\ref{fig1}(b) reaffirm the 2 
dimensionality of the spin dynamics in our 
sample\cite{la2dv}.  An extensive search along the (100)
and (101) directions in the basal plane yielded no 
incommensurate peaks, such as those found in Sr doped 
La$_2$CuO$_4$\cite{la2ch,L214_Sr,wakimasu}.
This means that the spin liquid in La$_2$Cu$_{0.94}$Li$_{0.06}$O$_4$ is 
composed of simple chess-board type dynamic antiferromagnetic spin clusters
in the CuO$_2$ plane, 
which have grown to substantial size below 100~K with the correlation length, 
$\xi \gg 42$~\AA, the inverse of the half-width-at-half-maximum 
of the in-plane peak. 
The correlation length is much longer than the mean
distance between Li dopants, 15~\AA.

The commensurate spatial magnetic structure in Li doped 
La$_2$CuO$_4$,
which is simpler than incommensurate structures observed
in Sr doped La$_2$CuO$_4$\cite{la2ch,wakimasu}, 
may be understood in light of the mobility of doped 
holes\cite{Li214,takagi}. 
Microscopic calculations have shown that the holes are loosely bound 
by the Li dopants\cite{h_haas}. Vortices associated
with the holes, which are a long-range and effective 
topological disturbance to
the 3D N\'{e}el order, preserve the commensurate ($\pi,\pi$) 
magnetic correlations\cite{h_haas,h_ctb}.
The incommensurate magnetic structures, on the other hand,
through either the ``stripe''\cite{stripe_jt} or the nesting Fermi 
surface\cite{nfs_si} mechanism, require more mobile holes such as those in
Sr doped La$_2$CuO$_4$\cite{takagi}. 
The effect of hole
mobility on spatial magnetic correlations has also been demonstrated in 
numerical simulation\cite{sq2dbym}.

\begin{figure}
\psfig{file=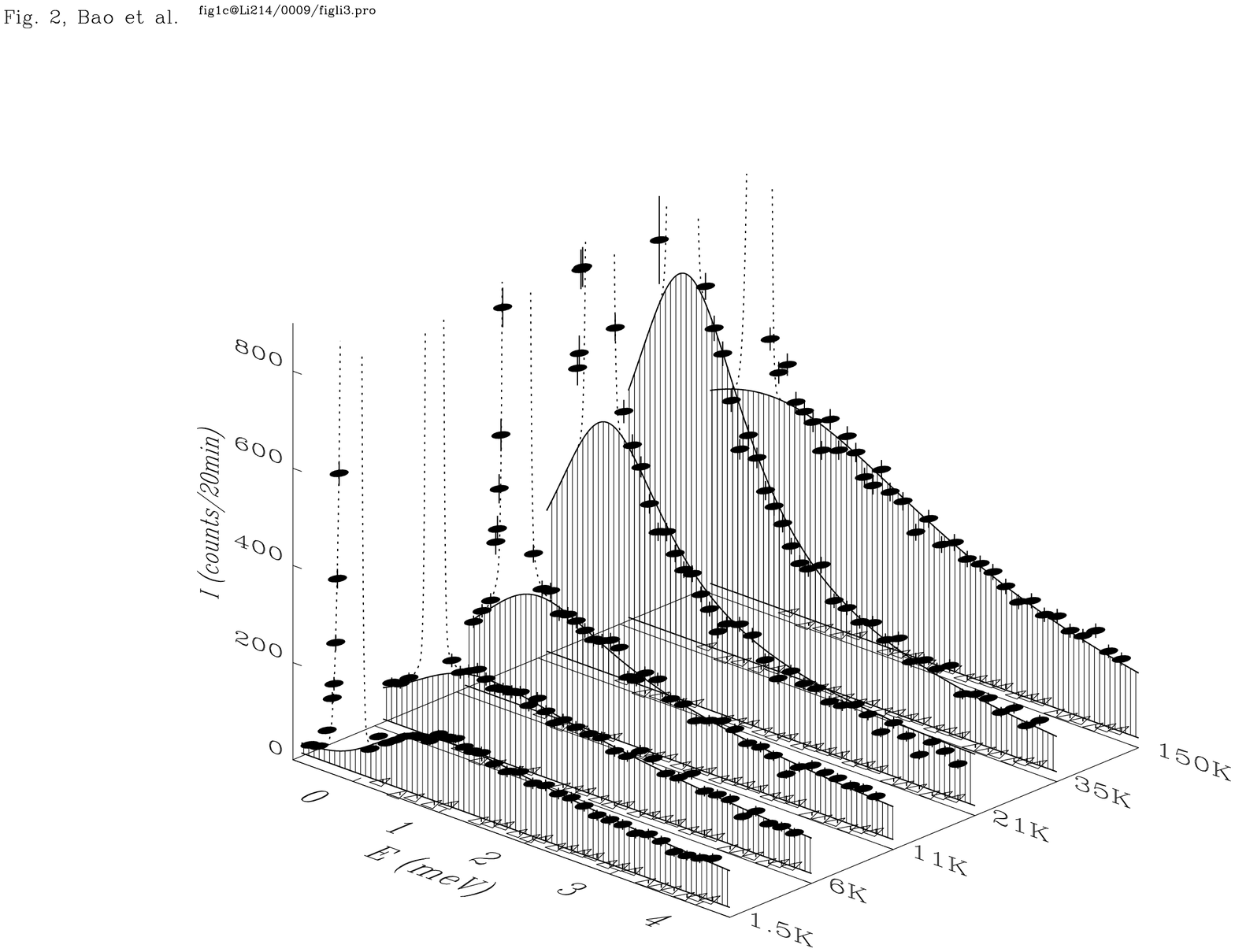,width=\columnwidth,angle=90,clip=}
\caption{
Const.-{\bf q}=(100) scans from 1.5 to 150~K.
The shaded area represents least-squares fit of 
inelastic magnetic intensity, $I(\omega)$,
to Eqs. (\ref{eq_I})-(\ref{eq_g}).
Triangles represent a flat background measured at (1.39,0,0).
}
\label{fig2}
\end{figure}

After delimiting the spatial part of spin dynamics 
in La$_2$Cu$_{0.94}$Li$_{0.06}$O$_4$, we now turn to the 
temporal dependence of
the 2D ($\pi,\pi$)-correlated spin liquid. Fig.~\ref{fig2}
shows energy scans at {\bf q}=(100)=($\pi,\pi$) at various temperatures.
Since magnetic intensity is sharply confined in {\bf q}-space in a rod 
passing through (100), as shown in Fig.~\ref{fig1},
the energy scan at (1.39,0,0) which is far away from the rod 
(triangles in Fig.~\ref{fig2}), 
offers a good measure of background.
In addition to the flat background, neutron scattering intensity 
in Fig.~\ref{fig2} consists of a convolution of the 
instrument resolution 
function with the dynamic magnetic structure factor 
$S(\omega,{\bf q})$
plus the elastic and incoherent peak at $\omega=0$\cite{bjbquasi}. 
The imaginary part of the generalized magnetic susceptibility
$\chi''(\omega,{\bf q})$ is related to the
dynamic magnetic structure factor $S(\omega,{\bf q})$ 
by the fluctuation-dissipation theorem\cite{neut_thry},
\begin{equation}
\chi''(\omega,{\bf q})= \pi \left(1-e^{-\hbar\omega/k_BT}\right)
	S(\omega,{\bf q}).   \label{eq_S}
\end{equation}
The fact that there is little change in measured {\bf q} widths in our energy
and temperature ranges (Fig.~\ref{fig1})
indicates that magnetic intensity in Fig.~\ref{fig2}
is proportional to the {\em local} dynamic magnetic structure factor,
$
I(\omega) \propto \int d{\bf q} S(\omega,{\bf q})   
$.
Therefore,
\begin{equation}
\chi''(\omega)\equiv \int d{\bf q} \chi''(\omega,{\bf q})=C
\left(1-e^{-\hbar\omega/k_BT}\right) I(\omega),   \label{eq_I}
\end{equation}
where $C$ is a normalization constant.
The commensurate spatial magnetic correlations and sharp in-plane peak
(long correlation length) make measurement of the local dynamic magnetic 
susceptibility, $\chi''(\omega)$, in La$_2$Cu$_{0.94}$Li$_{0.06}$O$_4$
considerably easier than in Sr or Ba doped La$_2$CuO$_4$, which
requires scans through the quartet of incommensurate satellite peaks 
at each $\omega$\cite{la2smha,la2keimer,la2gas}.

\begin{figure}
\psfig{file=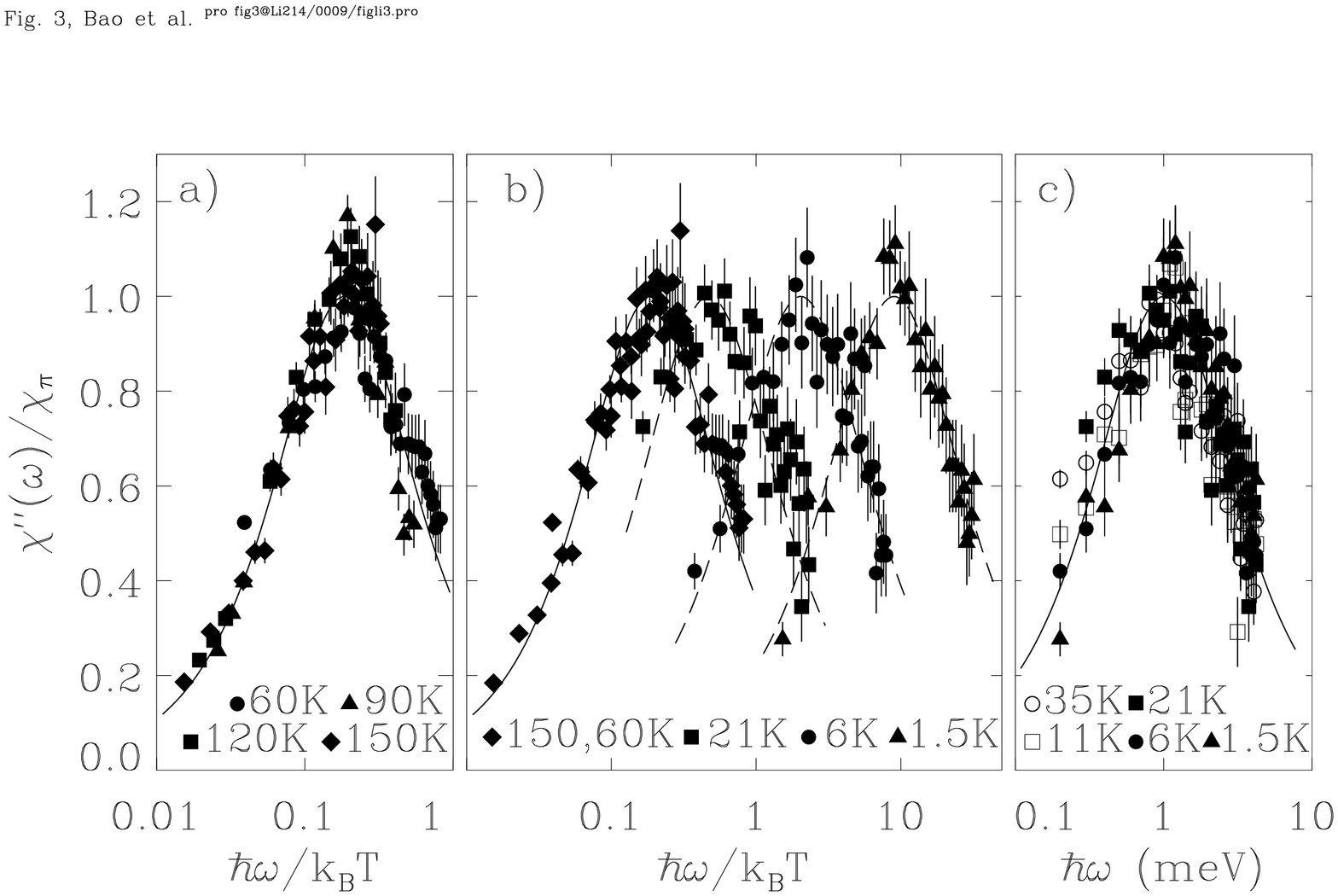,width=\columnwidth,angle=90,clip=}
\vskip -.7cm
\caption{(a) $\omega/T$ scaling is valid for La$_2$Cu$_{0.94}$Li$_{0.06}$O$_4$
in the high temperature QC regime. The solid line is the scaling 
function, Eq.~(\ref{eq_f}).
(b) $\omega/T$ scaling becomes invalid in the low temperature regime.
(c) A new scaling for the low temperature regime, with a constant energy scale
$\Gamma_0 \approx 1$~meV. The solid line is the scaling 
function, Eq.~(\ref{eq_g}).
}
\label{fig3}
\end{figure}

\begin{figure}
\psfig{file=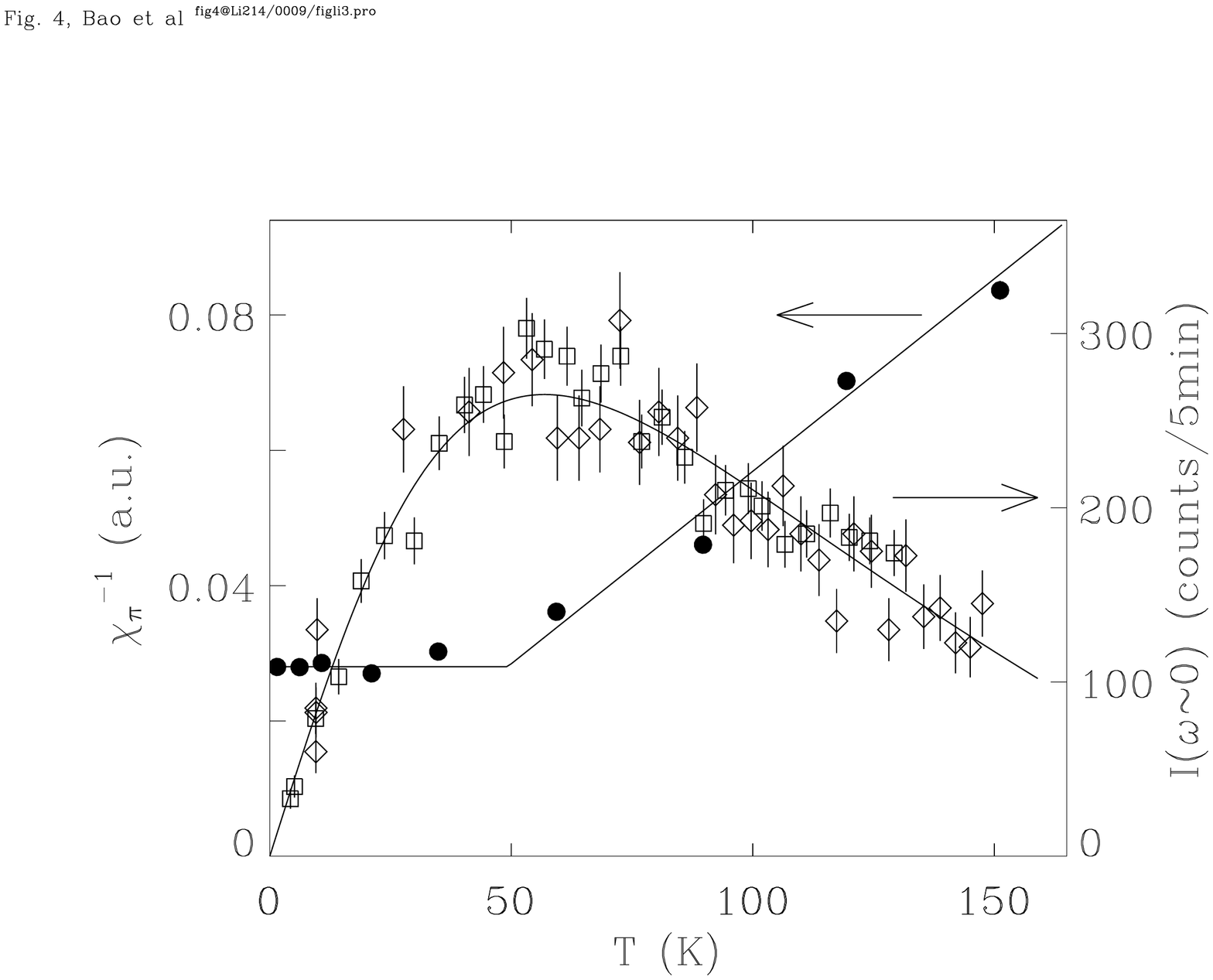,width=.9\columnwidth,angle=90,clip=}
\vskip -3ex
\caption{\label{fig4}
(left scale) Temperature dependence of inverse local magnetic susceptibility 
$\chi_{\bf \pi}^{-1}$ (circles) for La$_2$Cu$_{0.94}$Li$_{0.06}$O$_4$,
see Eq.~(\ref{eq_qc}) and (\ref{eq_qd}).
(right scale) Temperature dependence of magnetic intensity (open symbols)
measured at $\hbar\omega=0.2$~meV, $I(\omega\sim 0)$. 
The diamonds and squares were measured
during cooling and warming, respectively.
}
\end{figure}

$\chi''(\omega)$, obtained through Eq.~(\ref{eq_I}) from scans such 
as those in Fig.~\ref{fig2}, is shown as a normalized
function of the dimensionless scaling variable, $\hbar\omega$/k$_B T$, 
on a semi-logarithmic scale in Fig.~\ref{fig3}(a) and (b).
The scaling parameter $\chi_{\bf \pi}$ is shown in Fig.~\ref{fig4} 
(diamond) as a function of temperature.  $\omega/T$ 
scaling of $\chi''(\omega)$, 
which has been thoroughly investigated in pioneering
works on Ba and Sr doped La$_2$CuO$_4$\cite{la2gas,la2keimer,la2smha}, 
is also valid above 50~K for our Li doped sample La$_2$Cu$_{0.94}$Li$_{0.06}$O$_4$.
In Fig.~\ref{fig3}(a), the 150, 120, 90 and 60~K data are scaled onto a single 
universal curve 
\begin{equation}
\chi''(\omega)/\chi_{\bf \pi}=f(\hbar\omega/k_BT)  \label{eq_qc}
\end{equation}
with the scaling function
\begin{equation}
f(x)=\frac{0.18x}{0.18^2+x^2}, \label{eq_f}
\end{equation}
validating the quantum critical behavior at high temperatures. 
The factor, 0.18,
between the energy scale and thermal energy $k_BT$ is similar to
that in the insulating phase of Ba doped 
La$_2$CuO$_4$\cite{la2smha}.

Below 50~K, however, the $\omega/T$ scaling breaks down in
La$_2$Cu$_{0.94}$Li$_{0.06}$O$_4$, as 
clearly shown in Fig.~\ref{fig3}(b).   Data taken 
at 1.5, 6 and 21~K 
no longer fall on the solid $\omega/T$ scaling curve, Eq.\ (\ref{eq_f}),
valid above 50~K. 
Instead, a new scaling scheme describes the low temperature regime 
of the spin liquid.
Fig.~\ref{fig3}(c) shows the normalized $\chi''(\omega)$ 
as a function of energy for data taken at low temperatures. 
All data are described by
\begin{equation}
\chi''(\omega)/\chi_{\bf \pi}=g(\hbar\omega/\Gamma_0) \label{eq_qd}
\end{equation}
with the energy scale
$\Gamma_0\approx 1$~meV and the scaling function
\begin{equation}
g(x)=\frac{x}{1+x^2}. \label{eq_g}
\end{equation}
Notice also that $\chi_{\bf \pi}$ is basically
constant below 50~K (Fig.~\ref{fig4}), therefore, 
$\chi''(\omega)$ is essentially independent of temperature in the new 
low temperature regime.

The crossover from the high temperature QC regime to the new low 
temperature regime also manifests itself in low-energy 
magnetic neutron
scattering intensity. At high temperatures, from Eqs.~(\ref{eq_I}), 
(\ref{eq_qc}) and (\ref{eq_f}),
\begin{equation}
I(\omega \rightarrow 0) \propto \chi_{\bf \pi}\sim T^{-1}. \label{eq_ht}
\end{equation}
Therefore, magnetic intensity at the low energy limit
increases during cooling as the local magnetic susceptibility,
$\chi'(0) =\pi\,\chi_{\bf \pi}/2 $, does, consistent with common paramagnetic behavior.
Below the crossover temperature, from Eqs.~(\ref{eq_I}), (\ref{eq_qd}) and
(\ref{eq_g}) and noting the constant $\chi_{\bf \pi}$ in this regime,
\begin{equation}
I(\omega \rightarrow 0) \propto T, \label{eq_lt}
\end{equation}
which is markedly different from that in the QC regime, Eq.~(\ref{eq_ht}).
Circles in Fig.~\ref{fig4} show $I(\omega)$ measured at {\bf q}=(100)
and $\hbar\omega=0.2$~meV$\sim 0$, the lowest accessible 
energy outside the elastic 
incoherent peak 
(see Fig.\ref{fig2}).
The $I(\omega\sim 0)$ changes from an increasing to a decreasing
function of temperature as a consequence of the crossover in spin dynamics.
This is consistent with Eqs.~(\ref{eq_lt}) and (\ref{eq_ht}). 
The crossover temperature is around 50~K, where $I(\omega\sim 0)$
peaks and $\chi_{\bf \pi}$ begins to saturate. 
The crossover temperature is also consistent with $\Gamma_0/0.18k_B$,
obtained from equating the energy scales of the two regimes.

We have shown in this cold neutron inelastic study that spin dynamics 
in La$_2$Cu$_{0.94}$Li$_{0.06}$O$_4$
crosses over around 50~K from the quantum critical 
$\omega/T$ scaling to a new low 
temperature regime, characterized by a fixed energy scale and
by a temperature invariant spectrum $\chi''(\omega)$ in 
Eq.~(\ref{eq_qd}-\ref{eq_g}).
Now let us discuss the possible nature of the new regime.

In Sr or Li doped La$_2$CuO$_4$, a spin glass transition
occurs in a wide doping range at a similar temperature,
$T_{sf}\sim 10$~K\cite{muchn,fchou,Li214phs}. In
La$_2$Cu$_{0.94}$Li$_{0.06}$O$_4$,
$T_{sf}=8$~K\cite{Li214phs}. This raises the possibility
that the departure from the $\omega/T$ scaling at low
temperature might be caused by the spin glass transition.
However, the key spectral signature of a spin glass
transition is a zero energy scale as the time scale
becomes infinite\cite{sg_sms}. Instead of approaching
zero, the energy scale of
La$_2$Cu$_{0.94}$Li$_{0.06}$O$_4$ saturates at a finite
energy, $\Gamma\approx 1$~meV, at the crossover.  
Therefore, the crossover around 50~K is not related to the
spin glass transition.

A second possibility is to identify the low temperature regime with the
quantum disordered regime of the 2D Heisenberg 
antiferromagnet, which is 
characterized by a constant energy scale. As discussed in the introduction,
the finite energy scale is determined by the quantum long wavelength cutoff,
which depends on the distance from the quantum critical 
point, $y-y_c$\cite{2dheis,2dheiqd}. Physically, the distance can be tuned by
microscopic mechanisms, such as frustrating magnetic 
interactions\cite{2dheis,qpt_ss},
or vortex-like topological magnetic defects associated with 
holes\cite{h_haas,h_ctb}. In La$_2$Cu$_{0.94}$Li$_{0.06}$O$_4$,
the latter mechanism may apply. Note that the finite long wavelength cutoff 
due to hole-induced magnetic vortices here is different
from the upper cutoff length of percolating cluster on 
a square lattice, which is infinite for $y\lesssim 40$\%, 
as shown in a neutron scattering study 
on Zn and Mg doped La$_2$CuO$_4$\cite{opvmg}. 
 
In real materials, there are additional low temperature
phase transitions, which are caused by effects ignored in
theoretic models of the 2D quantum
antiferromagnet\cite{2dheis,2dheiqd,qpt_ss}. These
transitions include (1) a finite temperature N\'{e}el
transition due to weak interlayer magnetic interaction,
(2) a spin glass transition due to disorder unavoidable
with doping, (3) ``stripe'' formation\cite{stripe_jt} due
to the interaction between antiferromagnetic order and 
mobile holes, and (4) superconducting transition. 
Nevertheless, 2D quantum antiferromagnetism can be 
accessed in the
temperature window between $J/k_B$ and the undesired low
temperature phase transitions. The QC
regime\cite{la2gas,la2keimer,la2smha} as well as the
renormalized classical
regime\cite{la2dbir,scocmg,cftd,cftd_pc} of the 2D
antiferromagnetic spin liquid have been successfully
investigated experimentally in this window. Here, a
possible QC to QD crossover around 50~K falls also within
the $J$$\sim$1000~K and $T_{sf}$.

In effective low temperature theories based on the 2D 
quantum non-linear 
$\sigma$ model, 
which simulates the N\'{e}el order suppressing effect of
hole doping with frustration in magnetic interactions, 
the scaling function $g(x)$ is usually gapped\cite{2dheis,qpt_ss}.
When interaction between spin excitations and 
doped fermions is explicitly considered, however,
$g(x)$ becomes gapless\cite{2dheiz2,2dheisu}. 
The gapless $g(x)$ observed in La$_2$Cu$_{0.94}$Li$_{0.06}$O$_4$, 
Eq.~(\ref{eq_g}), may be an indicator of coupling between
holes and spin dynamics in doped cuprates.
More recently, quantum
orders are analyzed generally using projective symmetry 
groups\cite{q_wen}. Hundreds of symmetric spin liquids have been
constructed, which can be grouped into four classes.
Of the three stable classes in 2D, two are gapless.

In summary, the ($\pi,\pi$)-correlated dynamic spin clusters in hole-doped La$_2$Cu$_{0.94}$Li$_{0.06}$O$_4$ have developed to substantial size below
150~K.  The dynamics of such spin clusters cross over 
around 50~K from the quantum 
critical $\omega/T$ scaling to a new low temperature regime with a saturated
energy scale at $\Gamma_0\approx 1$~meV. The observed crossover possibly
corresponds to the theoretically expected quantum critical 
to quantum disordered 
crossover for 2D antiferromagnet.

We thank S.-H. Lee for hospitality and assistance at NIST;
P.C. Hammel, C. Broholm, A. Zheludev, L. Yu, E. Tosatti, S. Sachdev, 
A.V. Chubukov, F.C. Zhang, 
C.M. Varma, G. Aeppli, J. Haase, P. Carretta, N.J. Curro, 
E. Dagotto, A.V. Balatsky, Y. Bang, A. Abanov, D. Pines,
R. Heffner, P. Littlewood and A.P. Ramirez for useful discussions. 
SPINS at NIST is supported by NSF. Work at LANL is
supported by U.S. Department of Energy.


\end{document}